\documentclass[12pt]{article}

\usepackage{amsmath}
\usepackage{amssymb}
\usepackage{amsfonts}
\usepackage{lscape}

\begin{document}

\fontsize{12}{6mm}\selectfont
\setlength{\baselineskip}{2em}

$~$\\[.35in]
\newcommand{\dss}{\displaystyle}
\newcommand{\raro}{\rightarrow}
\newcommand{\be}{\begin{equation}}

\def\sech{\mbox{\rm sech}}
\thispagestyle{empty}

\begin{center}
{\Large\bf The Generalized Weierstrass System for } \\    [2mm]
{\Large\bf Nonconstant Mean Curvature Surfaces} \\    [2mm]
{\Large\bf and the Nonlinear Sigma Model}  \\  [2mm]
\end{center}

\vspace{1cm}
\begin{center}
{\bf Paul Bracken}                        \\
{\bf Department of Mathematics,} \\
{\bf University of Texas,} \\
{\bf Edinburg, TX  }  \\
{78541-2999}
\end{center}

\vspace{3cm}
\begin{abstract}
A study of the generalized Weierstrass system
which can be used to induce mean curvature surfaces
in three-dimensional Euclidean space is presented. 
A specific transformation is obtained which
reduces the initial system to a two-dimensional
Euclidean nonlinear sigma model. Some aspects
of integrability are discussed, in particular,
a connection with a version of the sinh-Gordon equation 
is established. Finally, some specific solutions
are given and a systematic way of calculating
multisoliton solutions is presented.
\end{abstract}

\newpage
{\bf I. Introduction.}
\par
The method which was first formulated by Weierstrass
and Enneper $[1]$ for determining minimal surfaces
embedded in three-dimensional Euclidean space
has recently been generalized by B. Konopelchenko $[2,3]$
and it has been the subject of considerable
further work $[4-7]$. One of the reasons for this is
that there are many physical applications of
minimal surfaces to such areas as integrable systems,
statistical mechanics and even string theory $[5,8]$.
In fact, a direct connection between certain
classes of mean curvature surface, namely
constant mean curvature, and a particular
integrable finite dimensional Hamiltonian system
has been established by Konopelchenko and
Taimanov $[3]$. Mean curvature plays a special
role among the characteristics of surfaces and
their dynamics in many problems that arise in
both physics and mathematics $[8]$. 
The case in which the mean curvature of
the surface is constant has been discussed $[4-5]$
at length. In particular, many properties of
such surfaces determined by Konopelchenko's
inducing prescription, such as the relationship
to the nonlinear two-dimensional sigma model,
integrability and Lax pair have been
determined and a B\"{a}cklund transformation
has been calculated as well $[6-7]$.

It is the purpose here to begin an investigation
of the case in which the mean curvature of the
surface is not constant, but is described by
a real-valued function which depends on two
independent variables.

This is useful for several reasons, in particular
for establishing some of the
mathematical properties of such surfaces
in general, beyond the case of constant mean curvature 
surfaces. There are many applications of mean
curvature surfaces and, of particular interest
here, we elaborate on applications to the areas
of quantum field theory, two-dimensional
gravity and string theory $[9-10]$. Classical string
theory can be regarded as a study of the geometry 
of subspaces immersed in higher dimensional
spaces of manifolds. The variational equations
of motion for the string or membrane
involve the mean curvature function.
These equations are usually obtained by
varying a particular functional,
such as the Willmore functional, which depends
explicitly on the mean curvature.
Once this is determined, the present approach
would provide a way of associating a particular
surface with the particular mean curvature function. 
The three-dimensional case 
has been discussed somewhat and may not be the
most significant in physical applications, but it is useful
to study and will hopefully lead the way to
generalizations to higher dimensional spaces. Thus, a
more sophisticated approach would involve the
calculation of the mean curvature from the
proposed equations of motion, which result
from the variation of some given action functional.
This results in a function that can be
employed in an inducing procedure such as the one
described here to generate a surface in a higher
dimensional space.
Generating the surface then amounts to solving
a system of first-order nonlinear partial
differential equations which depend on the mean 
curvature function for a set of functions.
These functions can be used to determine the
coordinates of a surface in Euclidean three space
corresponding to the given mean curvature function
by means of a particular inducing prescription.
One in particular which is relevant to the case of three space
will be presented here. A connection between the 
generalized Weierstrass system and a modified sigma
model system is established. This connection allows
solutions of the sigma model system to be related
to solutions of the generalized Weierstrass system
which can be used to produce surfaces. A set of
conservation laws relevant to the generalized
Weierstrass system is presented.
Integrability is reviewed
and a connection with the sinh-Gordon equation is made.
Finally, some specific solutions are determined for the 
sigma model system for specific mean curvature
and the corresponding solutions to the generalized Weierstrass
system are determined from these as well.

{\bf II. First Order System and Inducing Formulas}

The set of functions which are responsible
for determining the surface corresponding to 
a mean curvature function $H$ will be determined
from a system of coupled first order
differential equations. The following nonlinear 
Dirac-type system of differential equations
which determine two complex-valued functions
$\psi_1$ and $\psi_2$ corresponding to the
mean curvature function $H$ are used to
generate a surface immersed in $\mathbb R^3$.
This system is given by Konopelchenko {\bf [2-3]} in complex form as follows
$$
\begin{array}{ccc}
\partial \psi_1 = p H \psi_2,  &   &
\bar{\partial} \psi_2 = - p H \psi_1,   \\
   &    \\
\bar{\partial} \bar{\psi_1} = p H \bar{\psi}_{2},  &  &
\partial \bar{\psi}_{2} = - p H \bar{\psi}_{1},
\end{array}
\eqno(2.1)
$$
$$
p = | \psi_1 |^2 + |\psi_2|^2,
$$
where $\psi_1$ and $\psi_2$ are two complex functions
of the complex variables $(z, \bar{z})$. The bar denotes
complex conjugation in what follows and we define the
derivatives $\partial = \partial / \partial z$ and
$\bar{\partial} = \partial / \partial \bar{z}$. 
The function $H = H (z, \bar{z})$ denotes the 
mean curvature of the surface. The system (2.1) can be
considered as a variant of the original Weierstrass-Enneper
system $[1]$. Moreover, system (2.1) determines a
set of constant mean curvature surfaces which are
obtained by means of the following parametrization
$( z, \bar{z}) \rightarrow ( X_{1} ( z, \bar{z}),
X_{2} ( z , \bar{z} ), X_{3} ( z ,\bar{z}))$,
such that the $X_{j}$ can be determined explicitly
from the solutions $\psi_{i}$ to (2.1) by
evaluating the following set of integrals
$$
X_{1} + i X_{2} = 2 i \int_{z_0}^{z}
( \bar{\psi}_{1}^2 \, d z' - \psi_{2}^2 \, d \bar{z}' ),
$$
$$
X_{1} - i X_{2} = 2 i \int_{z_{0}}^z 
( \psi_{2}^2 \, d z' - \psi_{1}^2 \, d \bar{z}' ),
\eqno(2.2)
$$
$$
X_{3} =- 2 \int_{z_0}^z (  \bar{\psi}_1 \psi_2 \, d z'
+ \psi_1 \bar{\psi}_{2} \, d \bar{z}' ).
$$
The Gaussian curvature of the surface is then given by
$$
K = - \frac{\partial \bar{\partial} \log p}{p^2},
\eqno(2.3)
$$
where $p$ is given in (2.1) in terms of the $\psi_{i}$.
In (2.1), the function
$H ( z , \bar{z})$ is arbitrary, and so, for given $H (z, \bar{z})$,
a particular solution to (2.1) will generate a surface 
in Euclidean three space by means of equations (2.2).

{\bf III. Associated Second Order System.}

It is useful to subject the system described in
(2.1) to several transformations in order to
investigate its structure. In fact, it will be
shown that (2.1) has an associated second order
system whose solutions can be used to
determine surfaces by means of a transformation
linking these solutions to corresponding
solutions of (2.1). Begin by introducing the
new complex variable
$$
\rho = \frac{\psi_{1}}{\bar{\psi_{2}}},
\eqno(3.1)
$$
where $\psi_1$ and $\psi_{2}$ satisfy system
(2.1). To calculate $\partial \rho$, we
differentiate (3.1) and, upon using the relation
$p = |\psi_2|^2 ( 1 + |\rho|^2)$,
we obtain
$$
\partial \rho = \frac{\partial \psi_{1}}{\bar{\psi}_{2}}
- \frac{\psi_{1}}{\bar{\psi}_{2}^2} \partial \bar{\psi}_{2}
= \frac{p H}{\bar{\psi}_{2}^2} (|\psi_{1}|^2 + |\psi_{2}|^2)
= \frac{H p^2}{\bar{\psi}_{2}^2} = H \psi_{2}^2 ( 1 + |\rho|^2)^2.
$$
Solving this equation for $\psi_{2}$, we find that
$$
\psi_{2} = \epsilon \frac{(\partial \rho)^{1/2}}
{H^{1/2} ( 1 + |\rho|^2)}, \quad
\epsilon = \pm 1.
\eqno(3.2)
$$
From (3.1), it is easy to determine $\psi_1$ to be
$$
\psi_1 = \epsilon \frac{\rho ( \bar{\partial} \bar{\rho})^{1/2}}
{H^{1/2} ( 1 + |\rho|^2)},
\quad
\epsilon = \pm 1.
$$
This generates the following transformation from the
variable $\rho$ into the pair of variables $\psi_{i}$
provided that the $\psi_i$ satisfy (2.1). To summarize,
we have shown that
$$
\psi_{1} = \epsilon \rho \frac{(\bar{\partial} \bar{\rho})^{1/2}}
{H^{1/2} ( 1 + |\rho|^2)},
\qquad
\psi_{2} = \epsilon \frac{(\partial \rho)^{1/2}}
{H^{1/2} ( 1 + |\rho|^2)},
\qquad
\epsilon = \pm 1.
\eqno(3.3)
$$

{\em Proposition 1. } If $\psi_1 $ and $\psi_{2}$ are
solutions of system (1.1), then the function $\rho$
defined by (3.1) and related to $\psi_1$ and $\psi_2$
by (3.3) is a solution of the following
second order system
$$
\partial \bar{\partial} \rho
- \frac{2 \bar{\rho}}{1 + |\rho|^2}
\partial \rho \, \bar{\partial} \rho
= \bar{\partial} ( \ln H) \partial \rho,
\qquad
\bar{\partial} \partial \bar{\rho}
- \frac{2 \rho}{1 + |\rho|^2}
\bar{\partial} \bar{\rho} \partial \bar{\rho}
= \partial (\ln H) \bar{\partial} \bar{\rho}.
\eqno(3.4)
$$

{\em Proof:} Differentiating the equation
$\partial \rho = H p^2 / \bar{\psi}_{2}^2$
with respect to $\bar{z}$, we obtain that
$$
\bar{\partial} \partial \rho = \frac{\bar{\partial} H \, p^2}
{\bar{\psi}_{2}^2} + 2 H \frac{p \bar{\partial} p}{\bar{\psi}_{2}^2}
- 2 H \frac{p^2}{\bar{\psi}_{2}^3} ( \bar{\partial} \bar{\psi}_{2})
$$
$$
= \frac{H p^2}{\bar{\psi}_{2}^2} [ \bar{\partial} (\ln H)
+ \frac{2}{p \bar{\psi}_{2}} ( \bar{\psi}_{1}
\bar{\psi}_{2} (\bar{\partial} \psi_{1}) + |\psi_{2}|^2
(\bar{\partial} \bar{\psi}_{2}) - |\psi_{1}|^2
\bar{\partial} \bar{\psi}_{1} - |\psi_{2}|^2
\bar{\partial} \bar{\psi}_{2} ) ]
$$
$$
= \frac{p H}{\bar{\psi}_{2}^3} [ p \bar{\psi}_2
\bar{\partial} ( \ln H) + 2 ( \bar{\psi}_{1} \bar{\psi}_{2}
( \bar{\partial} \psi_{1}) - |\psi_{1}|^2
( \bar{\partial} \bar{\psi}_{2}))].
$$
Substituting the derivatives into the left
hand side of (3.4), we have
$$
\partial \bar{\partial} \rho
- \frac{ 2 \bar{\rho}}{1 + |\rho|^2} 
\partial \rho \bar{\partial} \rho
$$
$$
= \frac{p H}{\bar{\psi}_{2}^3} 
[ p \bar{\psi}_2 \bar{\partial} ( \ln H)
+ 2 \bar{\psi}_1 \bar{\psi}_{2}
( \bar{\partial} \psi_1) - 2 |\psi_1|^2
( \bar{\partial} \bar{\psi}_2)]
- \frac{2 ( \bar{\psi}_{1}/ \psi_2) |\psi_2|^2}{|\psi_{1}|^2 
+ |\psi_{2}|^2} \frac{H p^2}{\bar{\psi}_{2}^4}
(\bar{\psi}_{2} \bar{\partial} \psi_1 - \psi_1 \bar{\partial}
\bar{\psi}_{2})
$$
$$
= \frac{pH}{\bar{\psi}_{2}^3} [ p \bar{\psi}_{2}
\bar{\partial} (\ln H)
+ 2 \bar{\psi}_{1} \bar{\psi}_{2} ( \bar{\partial} 
\psi_{1}) - 2 |\psi_{1}|^2 \bar{\partial} \bar{\psi}_{2}
- 2 \bar{\psi}_{1} \bar{\psi}_{2} \bar{\partial}
\psi_{1} + 2 |\psi_1|^2 \bar{\partial} \bar{\psi}_{2}]
$$
$$
= \frac{p^2 H}{\bar{\psi}_{2}^2} \bar{\partial}
( \ln H) = (\bar{\partial} \ln H) \partial \rho.
$$
The second conjugate equation in (3.4) can be obtained
in the same way. QED

There is a converse to Proposition 1, and this
is formulated in the following statement.

{\em Proposition 2.} If $\rho$ is a solution to
system (3.4), then the functions $\psi_{1}$
and $\psi_{2}$ defined in (3.3) in terms of
$\rho$ satisfy (2.1).

{\em Proof:} Differentiate $\psi_1$ given by
(3.3) with respect to $z$ and eliminate the second derivative
$\partial \bar{\partial} \bar{\rho}$ using
(3.4) to obtain
$$
\partial \psi_1 = \epsilon \partial \rho
\frac{(\bar{\partial} \bar{\rho})^{1/2}}
{H^{1/2} ( 1 + |\rho|^2)}
+ \frac{1}{2} \frac{\epsilon \rho}{H^{1/2} 
(1 + |\rho|^2)} (\frac{2 \rho}{1 +|\rho|^2}
\partial \bar{\rho} + \partial (\ln H) )
( \bar{\partial} \bar{\rho})^{1/2}
$$
$$
-\frac{1}{2} \epsilon \rho
\frac{( \bar{\partial} \bar{\rho})^{1/2}
\partial ( \ln H)}
{H^{1/2} ( 1 + |\rho|^2)}
- \epsilon \rho \frac{(\bar{\partial} \bar{\rho})^{1/2}}
{H^{1/2} ( 1 + |\rho|^2)^2} (\partial \rho \bar{\rho}
+ \partial \bar{\rho} \rho)
$$
$$
= \epsilon \partial \rho (1 + |\rho|^2)
\frac{(\bar{\partial} \bar{\rho})^{1/2}}
{H^{1/2} ( 1 + |\rho|^2)^2} 
- \epsilon |\rho|^2
\partial \rho \,
\frac{(\bar{\partial} \bar{\rho})^{1/2}}
{H^{1/2} ( 1 + |\rho|^2)^2}
= H (z, \bar{z}) p \psi_{2}.
$$
The other equation in (2.1) is developed by
differentiating $\psi_2$ in (3.3) with
respect to $\bar{z}$ and simplifying
as in the case presented above.

There are several conservation laws as well
associated with system (2.1) which are important
and need to be discussed. One reason is that these results
will imply the path independence of the integrals which
generate the coordinate functions of the surface
given in (2.2). It can be noted that the system 
(3.4) is invariant under discrete transformations
generated by the reflections
$$
Z_1 : \quad z \rightarrow z, \qquad
\bar{z} \rightarrow - \bar{z}, \qquad
\rho \rightarrow \rho,   \qquad
\bar{\rho} \rightarrow \bar{\rho},
$$
$$
Z_2 : \quad z \rightarrow z,  \qquad
\bar{z} \rightarrow \bar{z},  \qquad
\rho \rightarrow - \rho,  \qquad
\bar{\rho} \rightarrow - \bar{\rho},
$$
$$
Z_3 : \quad z \rightarrow -z \qquad
\bar{z} \rightarrow \bar{z},  \qquad
\rho \rightarrow - \rho,    \qquad
\bar{\rho} \rightarrow - \bar{\rho},
$$
their complex conjugates as well as the inversion 
$$
I : z \rightarrow z, \qquad  \bar{z} \rightarrow \bar{z} \qquad
\rho \rightarrow 1/ \rho, \qquad \bar{\rho}
\rightarrow 1 / \bar{\rho}.
$$

There is the conservation of a potential
function with respect to (2.1), namely,
$$
\partial (\psi_{1}^2) + \bar{\partial} (\psi_{2}^2)
= 2 \psi_{1} \partial \psi_{1} + 2 \psi_{2}
\bar{\partial} \psi_{2} = 2 \psi_{1} p H \psi_{2}
+ 2 \psi_2 ( -p H \psi_{1}) = 0,
\eqno(3.5)
$$
The conjugate of this equation holds as well. Moreover,
there is a further conservation relation
$$
\partial ( \psi_{1} \bar{\psi}_{2}) + 
\bar{\partial} ( \bar{\psi}_{1} \psi_{2} ) = 0.
\eqno(3.6)
$$
Next, let us define the quantity
$$
J = \bar{\psi}_{1} \partial \psi_{2} - \psi_{2}
\partial \bar{\psi}_{1}.
\eqno(3.7)
$$
Differentiating $J$ in (3.7) with respect to
$\bar{z}$, we find that
$$
\bar{\partial} J = \bar{\partial} \bar{\psi}_{1} 
\partial \psi_{2} + \bar{\psi}_{1} \bar{\partial} 
\partial \psi_{2} - \bar{\partial} \psi_2
\partial \bar{\psi}_{1} - \psi_{2} \bar{\partial}
\partial \bar{\psi}_{1}
$$
$$
= p H \bar{\psi}_{2} \partial \psi_2 + \bar{\psi}_1
\partial (-pH \psi_1) + pH \psi_1 \partial \bar{\psi}_1
- \psi_2 \partial (p H \bar{\psi}_{2})
$$
$$
= p H \bar{\psi}_{2} \partial \psi_2 - \bar{\psi}_{1}
( \partial H \, p \psi_{1} + H \partial p \,
\psi_1
+ p H \partial \psi_1) + p H \psi_1 \partial \bar{\psi}_1
$$
$$
- \psi_2 ( \partial H \, p \bar{\psi}_{2} + H
\partial p \bar{\psi}_{2} + p H \partial \bar{\psi}_{2})
$$
$$
= p H ( \partial p) - p^2 \partial H - p H \partial p
- p H \bar{\psi}_{1} (p H \psi_2) - p H \psi_2
( - p \bar{\psi}_{1}) = - p^2 ( \partial H).
\eqno(3.8)
$$
Clearly, if $H$ is constant then $\bar{\partial} J = 0$,
in which case $J$ is a conserved quantity, or current. 
Nonconstant $H$ breaks the conservation of this
current and $J$ given by (3.7) is no longer conserved. However,
using the result (3.8), we can modify $J$ to a new form 
which is conserved under differentiation by $\bar{z}$.

{\em Proposition 3.} The quantity defined in terms of $J$
from (3.7) in the form
$$
{\cal J} = J + \int_{\bar{z}_0}^{\bar{z}} p^2 (z, \tau)
\partial H ( z, \tau ) \, d \tau,
\eqno(3.9)
$$
where $\bar{z}_{0}$ is fixed in $\mathbb C$ is
conserved under differentiation with respect
to $\bar{z}$.

{\em Proof:} To prove this, differentiate (3.9) 
with respect to $\bar{z}$ and use (3.8) to obtain
$$
\bar{\partial} {\cal J} =
\bar{\partial} J + p^2 ( z, \bar{z}) \partial H
(z, \bar{z}) = - p^2 (\partial H) + p^2 (\partial H) =0.
$$
QED

Therefore, ${\cal J}$ defined by (3.9) is conserved
with respect to differentiation with respect to
$\bar{z}$. We can represent the integral operator
used in (3.9) by $\bar{\partial}^{-1}$ and write
the conserved quantity ${\cal J}$ in the following form
$$
{\cal J} = J + \bar{\partial}^{-1} (p^2 H),
\eqno(3.10)
$$
where in this notation $\bar{\partial} \bar{\partial}^{-1}
= \bar{\partial}^{-1} \bar{\partial} = {\bf 1}$.

\newpage
{\bf IV. Integrability Properties.}

It has been shown $[4]$ that the generalized 
Weierstrass system has the property of integrability
when the mean curvature is constant. Now we are interested
in conditions under which system (2.1) becomes a
completely integrable system. Employing the
conditional symmetry method $[11]$, we look for conditions
necessary for solvability of a class of equations
(3.4) which admit compatible first order differential
constraints. We consider here the simplest case 
where the differential constraints are based on an
$sl (2, \mathbb C)$ representation. Assume then that
they take the form of coupled Riccati equations
with nonconstant coefficients and
their complex conjugates, 
$$
\begin{array}{c}
\partial \rho = A^0_{1} ( z, \bar{z}) + A_{1}^1 ( z, \bar{z}) \rho
+ A_{1}^2 ( z, \bar{z}) \rho^2,  \\
  \\
\bar{\partial} \rho = A_{2}^0 ( z, \bar{z}) + A_{2}^1 (z, \bar{z})
\rho + A_{2}^2 ( z, \bar{z}) \rho^2.  \\
\end{array}
\eqno(4.1)
$$
To ensure that the pair of equations in (4.1)
satisfies a compatibility condition, it must be
required that the coefficient functions satisfy
the following system of zero curvature conditions
$$
\begin{array}{c}
\bar{\partial} A^0_1 - \partial A^0_{2} + A_{1}^1 A^0_{2}
- A^1_{2} A^0_{1} = 0,  \\
   \\
\bar{\partial} A^1_1 - \partial A^1_2 +2 A^2_1 A^0_2
-2 A^2_2 A^0_1 = 0,  \\
   \\
\bar{\partial} A^2_1 - \partial A^2_2 + A^2_1 A_2^1 - A^1_1 A^2_2 =0.
\end{array}
\eqno(4.2)
$$
We look for conditions on the function $H$
which ensure that the overdetermined system composed
of (3.4), differential constraints (4.1) and
conditions (4.2) are in involution. These 
involutivity conditions give the specific 
differential restrictions on the function $H$
$$
\bar{\partial} \partial ( \frac{1}{H} ) = 0.
\eqno(4.3)
$$
A general solution of equation (4.3) is given by
$$
H = \frac{1}{Q (z) + Q ( \bar{z}) },
\eqno(4.4)
$$
where $Q (z)$ is an arbitrary real valued
function, and (4.4) is written in this form so
that $H$ goes into itself under conjugation.
In this case, system (2.1) can be written as
$$
\begin{array}{cc}
\partial \psi_1 = \dss \frac{p}{Q(z) + Q (\bar{z})} \psi_2  &
\dss \bar{\partial} \bar{\psi}_{1} = \frac{p}{Q(z) + Q (\bar{z})} \bar{\psi}_2, \\
   &    \\
\dss \bar{\partial} \psi_2 = - \dss \frac{p}{Q(z) + Q (\bar{z})} \psi_1,  &
\dss \partial \bar{\psi}_{2} = - \dss \frac{p}
{Q (z) + Q ( \bar{z})} \bar{\psi}_{1}.
\end{array}
\eqno(4.5)
$$
Another way to write the solution to (4.3)
is $H = (h(z) + \bar{h} (\bar{z}))^{-1}$,
where $h(z)$ is a holomorphic function.
Beginning with the definition of $p$ in (2.1),
then modulo system (2.1), it is straightforward
to show that
$$
\partial p = \psi_1 \partial \bar{\psi}_{1} +
\bar{\psi}_{2} \partial \psi_2,
\quad
\bar{\partial} p = \bar{\psi}_{1} \bar{\partial}
\psi_1 + \psi_2 \bar{\partial} \bar{\psi}_{2}.
\eqno(4.6)
$$
Differentiating $\partial p$ in (4.1) with respect
to $\bar{z}$, we obtain an expression for $\bar{\partial}
\partial p$ as follows
$$
\bar{\partial} \partial p = \bar{\partial} \psi_1
\partial \bar{\psi}_1 + \psi_1 \bar{\partial} \partial \bar{\psi}_{1}
+ \bar{\partial} \partial \psi_2 \bar{\psi}_2 + \partial
\psi_2 \bar{\partial} \bar{\psi}_{2}
$$
$$
= \bar{\partial} \psi_1 \partial \bar{\psi}_{1} +
\psi_1 ( \partial p H \bar{\psi}_{2} + p \partial H
\bar{\psi}_{2} + p H \partial \bar{\psi}_{2}) - \bar{\psi}_{2}
(\partial p H \psi_1 
+ p \partial H \psi_1 + p H \partial \psi_1) + 
\partial \psi_2 \bar{\partial} \bar{\psi}_{2}
$$
$$
= \bar{\partial} \psi_1 \partial \bar{\psi}_{1}
+ \partial \psi_2 \bar{\partial} \bar{\psi}_{2} - p^3 H^2.
\eqno(4.7)
$$
It has been shown that $p$ satisfies a modified
sinh-Gordon equation when $H$ is constant $[5]$. Here,
it will be shown that, with respect to system (2.1),
$p$ satisfies essentially an identical equation.

{\em Proposition 4.} Let $p$ be defined 
by (2.1) and $J$ defined
by (3.7), then $p$ satisfies a second order differential
equation which involves $p$, $J$ and the mean
curvature $H$ and is given by
$$
\partial \bar{\partial} \ln p = \frac{|J|^2}{p^2}
- p^2 H^2.
\eqno(4.8)
$$

{\em Proof:} The derivative $\partial \bar{\partial} 
\ln p$ is expanded in the form
$$
\partial \bar{\partial} \ln p
 = \frac{1}{p^2} ( p \partial \bar{\partial} p
- \partial p \, \bar{\partial} p).
$$
The derivatives on the right hand side have been
evaluated and are substituted from (4.1) and (4.2).
We obtain that
$$
\partial \bar{\partial} \ln p = \frac{1}{p^2}
( p ( \bar{\partial} \psi_1 \partial \bar{\psi}_{1}
+ \partial \psi_2 \bar{\partial} \bar{\psi}_{2}) - p^4 H^2
- ( \psi_1 \partial \bar{\psi}_{1} + \bar{\psi}_2
\partial \psi_2)( \bar{\psi}_{1} \bar{\partial} \psi_1
+ \psi_2 \bar{\partial} \bar{\psi}_{2}))
$$
$$
= \frac{1}{p^2} ( |\psi_1|^2 \partial \psi_2 
\bar{\partial} \bar{\psi}_{2} + |\psi_{2}|^2
\bar{\partial} \psi_2 \partial \bar{\psi}_{1}
- \bar{\psi}_{1} \bar{\psi}_{2} \bar{\partial} \psi_1
\partial \psi_2 - \psi_1 \psi_2 \partial
\bar{\psi}_{1} \bar{\partial} \bar{\psi}_{2} - p^4 H^2).
\eqno(4.9)
$$
Using the definition of $J$ given in (3.7), it
follows immediately that
$$
|J|^2 = | \psi_1|^2 \partial \psi_2
\bar{\partial} \bar{\psi}_{2} - \psi_1 \psi_2
\partial \bar{\psi}_{1} \bar{\partial} \bar{\psi}_{2}
- \bar{\psi}_{1} \bar{\psi}_{2} \partial \psi_2
\bar{\partial} \psi_1 + |\psi_{2}|^2
\partial \bar{\psi}_{1} \bar{\partial} \psi_1.
$$
Comparing this expression for $|J|^2$ with the first four
terms in the bracket in (4.9), it follows
that $p$ satisfies the equation
$$
p^2 \partial \bar{\partial} \ln p = |J|^2 - p^4 H^2.
$$
Dividing both sides of this equation by $p^2$,
we immediately obtain (4.8).

It has been shown $[4]$ that when $H$ is
constant, there is a connection between the
time independent Landau-Lifshitz equation
which takes the form
$$
[ S , \partial \bar{\partial} S ] = 0,
\eqno(4.10)
$$
and the two-dimensional Euclidean nonlinear
sigma model $[12-14]$. The matrix $S$ will be
referred to as the spin matrix. In terms
of $\rho$ the sigma model variable, 
the matrix $S$ takes the form
$$
S = \frac{1}{1 + |\rho|^2} \left(
\begin{array}{cc}
1 - |\rho|^2    &  2 \bar{\rho}   \\
2 \rho  &  -  1 + |\rho|^2   \\
\end{array}   \right).
\eqno(4.11)
$$
The sigma model system (3.4) for the case in which $\rho$
is constant can be written in the form
$$
\partial \bar{\partial} \rho
- \frac{2 \bar{\rho}}{1 + |\rho|^2} 
\partial \rho \bar{\partial} \rho = 0,
\quad
\bar{\partial} \partial \bar{\rho}
- \frac{ 2 \rho}{1 + |\rho|^2} \bar{\partial}
\bar{\rho} \, \partial \bar{\rho} = 0,
\eqno(4.12)
$$
Define $f$ and $\bar{f}$ to be the $\rho$ dependent
factors on the left hand side of each respective equation in (4.12).
Then when $H$ is constant, from (4.12) the sigma model equations 
can be written in the form $f=0$, $\bar{f} = 0$. 
The matrix generated by (4.10) in terms of $f$ and $\bar{f}$ is
given explicitly by
$$
[ S, \partial \bar{\partial} S] = \frac{4}{(1 + |\rho|^2)^2} \left(
\begin{array}{cc}
\bar{\rho} f - \rho \bar{f}  & 
\bar{\rho}^2 f - \bar{f}  \\
  &     \\
-(f - \rho^2 \bar{f})   &  
-(\bar{\rho} f - \rho \bar{f})   \\
\end{array}
\right).
$$
We can summarize this in the following Proposition.

{\em Proposition 5.} If $\rho$ is a solution of the
nonlinear sigma model system (4.12),
then the spin matrix $S$ defined by
(4.11) is a solution of the Landau-Lifshitz
equation (4.10).

Of course, the spin matrix $S$ can be written
in terms of $\psi_1$ and $\psi_{2}$, and
Proposition 5 is altered to require that
$\psi_{1}$ and $\psi_2$ are solutions of the
generalized Weierstrass system, (2.1) with $H$ constant.

At this point, we would like to adapt this result
to the case in which $H$ is not constant.
To this end, we consider a nonhomogeneous or
deformed analogue of (4.10) given by
$$
 [ S , \partial \bar{\partial} S] + {\cal R} {\cal H} = 0.
\eqno(4.13)
$$
and the matrices ${\cal R}$ and ${\cal H}$ are given by
$$
{\cal R} = \frac{4}{( 1 + |\rho|^2)^2} \left(
\begin{array}{cc}
- \bar{\rho} \partial \rho  & \rho 
\bar{\partial} \bar{\rho}   \\
\partial \rho  &  - \rho^2 \bar{\partial} \bar{\rho}  \\
\end{array}
\right),
\quad
{\cal H} = \left(
\begin{array}{cc}
\bar{\partial} \ln H  &  \bar{\rho} \bar{\partial} \ln H  \\
  &     \\
\partial \ln H   &  \dss \frac{1}{\rho} \partial \ln H   \\
\end{array}
\right).
\eqno(4.14)
$$
where the matrix ${\cal R}$ depends only on the $\rho$ variable.
It is straightforward to calculate the product
${\cal R} {\cal H}$ and so (4.13) takes the form
$$
[ S, \partial \bar{\partial} S] + {\cal R} {\cal H} =
\frac{4}{( 1 + |\rho|^2)^2}
$$
$$
\cdot
\left(
\begin{array}{cc}
\bar{\rho} (f - ( \bar{\partial} \ln H) \partial \rho )
- \rho ( \bar{f} - ( \partial \ln H) \bar{\partial} \bar{\rho}) &
\bar{\rho}^2 ( f - ( \bar{\partial} \ln H) \partial \rho)
- ( \bar{f} - ( \partial \ln H) \bar{\partial} \bar{\rho}))  \\
       &     \\
- ( f - (\bar{\partial} \ln H) \partial \rho -
\rho^2 ( \bar{f} - ( \partial \ln H) \bar{\partial} \bar{\rho})) &
- ( \bar{\rho} ( f - ( \bar{\partial} \ln H) \bar{\partial} \rho)
- \rho ( \bar{f} - ( \partial \ln H) \bar{\partial} \bar{\rho}))  \\
\end{array}
\right)
\eqno(4.15)
$$

This gives the following generalization of Proposition 5.

{\em Proposition 6.} If $\rho$ is a solution of nonlinear sigma
model equations (3.4), and the matrices ${\cal R}$ and ${\cal H}$ in (4.13)
are defined by (4.14), then the spin matrix $S$ given in
(4.11) is a solution of the nonhomogeneous Landau-Lifshitz equation
(4.13) modulo (3.4).

In analogy with the constant mean curvature case,
there is a set of conditions which permit the system (2.1)
to become a linear decoupled system of equations which have
nonconstant coefficients.

{\em Proposition 7:} If the functions $\psi_1$ and $\psi_2$
satisfy the overdetermined system composed of the equations
of motion (2.1) and differential constraints
$$
\bar{\psi}_{1} \bar{\partial} \psi_1 + \psi_2 \bar{\partial} \bar{\psi}_{2} = 0,
\qquad
\bar{\psi}_{2} \partial \psi_2 + \psi_1 \partial \bar{\psi}_1 = 0,
\eqno(4.16)
$$
then the overdetermined system is equivalent to a linear
system with nonconstant coefficients of the form
$$
\bar{\partial} \partial \psi_1 - ( \bar{\partial} \ln H) 
\partial \psi_1 + p_{0}^2 H^2 \psi_1 =0,
\quad
\partial \bar{\partial} \psi_2 - (\partial \ln H) 
\bar{\partial} \psi_2 + p_{0}^2 H^2 \psi_2 = 0,
\eqno(4.17)
$$
and their respective conjugate equations, where
$| \psi_1|^2 + |\psi_2|^2 = p_{0}^2 \in \mathbb R$.

{\em Proof:} Making use of (4.6) and the conditions
(4.16), we obtain that the derivatives of $p$
vanish, $\partial p = 0$ and $\bar{\partial} p = 0$.
This means that if (4.16) hold, then $p$ is a real
constant $p_{0}$. Thus, $|\psi_1|^2 + |\psi_2|^2 = p_{0}$
is a conserved quantity. Differentiating system
(2.1) and replacing the known derivatives, we
obtain the second order system (4.17).

{\bf V. Determination of Specific Solutions.}

In this section, we would like to determine
specific examples of solutions for system (2.1). 
To this end, system (3.4) can be exploited to
determine functions $\rho$ corresponding to the given function
$H ( z, \bar{z})$, which describes the mean
curvature of the surface. Once these functions
$\rho$ have been obtained, equations
(3.3) are used to determine $\psi_1$ and
$\psi_2$ from $\rho$, and (2.2) then produces the
coordinates of the associated surface
of mean curvature $H$.

A specific class of solution to (3.4)
has been extensively investigated $[5,6]$ 
in the case in which the mean curvature is constant. 
This concerns the case in which $\rho$ 
satisfies $| \rho |^2 = 1$, that is, solutions
which are unimodular. Such solutions lead in a
straightforward way to the construction of multisoliton
solutions. It is worth showing here that such
solutions exist only for the case in which
the mean curvature $H$ is constant.

{\em Proposition 8.} Suppose $\rho$ is a
solution of the sigma model system (3.4)
which satisfies the condition $| \rho |^2 = 1$.
Then it follows that the mean curvature 
function $H$ in (3.4) is a constant.

{\em Proof:} Substituting $| \rho|^2 =1$ into system 
(3.4) it reduces to
$$
\bar{\partial} \ln \rho - \bar{\rho} \bar{\partial} 
\rho = \bar{\partial} \ln H,    \qquad
\partial \ln \bar{\rho} - \rho \partial \bar{\rho}
= \partial \ln H. 
\eqno(5.1)
$$
Integrating each of the equations in (5.1) in turn
with respect to $\bar{z}$ and $z$ respectively,
we obtain
$$
\ln \rho - \int \bar{\rho} \bar{\partial} \rho \,
d \bar{z} + f(z) = \ln H,
\qquad
\ln \bar{\rho} - \int \rho \partial \bar{\rho} \, d z 
+ g(z) = \ln H.
\eqno(5.2)
$$
Therefore $H$ must be given equivalently in the form
$$
H = \rho e^{- \int \bar{\rho}
\bar{\partial} \rho \, d \bar{z} + f (z)},
\qquad
H = \bar{\rho} e^{- \int \rho \partial \bar{\rho} \, dz + g(z)}
\eqno(5.3)
$$
Now substitute $\bar{\rho} = 1 / \rho$ 
into (5.3) and simplify to obtain
$$
H = \rho e^{- \int \bar{\partial} \ln \rho \, d \bar{z}
+ f(z)}
= \rho e^{- \ln \rho + f(z)} = e^{f(z)},
\qquad
H = \bar{\rho} e^{\int \partial \ln \rho \, dz + g( \bar{z})}
= \bar{\rho} e^{\ln \rho + g (\bar{z})}= e^{g ( \bar{z})}.
$$
Since these have to give the same real valued function $H$,
$f$ and $g$ must be equal to the same real constant
$f (z) =g ( \bar{z}) = c$, which implies that $H$ is constant.

When the mean curvature $H$ is constant, 
the following Proposition gives a way of constructing
multisoliton solutions from individual unimodular solutions {\bf [5]}.

{\em Proposition 9.} If $\rho_1$ and $\rho_2$
are unimodular and as well satisfy the sigma model system (4.12)
for constant $H$, then the product of the two functions
$\rho = \rho_1 \cdot \rho_2$ also satisfies the sigma model
system.

{\em Proposition 10.} Let $H$ be a 
mean curvature function which is identically constant
and $\varphi$ a function which
satisfies the pair of relations
$$
\partial \varphi = \ln ( \partial \ln \rho),
\qquad
\bar{\partial} \varphi = \ln H,
\eqno(5.4)
$$
where $\rho$ is unimodular. The compatibility condition
for (5.4) is equivalent to the sigma model system
(4.12).

{\em Proof.} Differentiating the first equation
in (5.4) with respect to $\bar{z}$ and the second
with respect to $z$, we obtain the results
$$
\bar{\partial} \partial \varphi =
\frac{\bar{\partial} \partial \rho}{\partial \rho} -
\frac{\bar{\partial} \rho}{\rho},
\qquad
\partial \bar{\partial} \varphi =0.
$$
Equating these, multiplying both sides by $\partial \rho$
and applying the constraint $ \bar{\rho} = 1/ \rho$, 
it is found that $\rho$ 
satisfies the first equation in (4.12). Using the complex conjugates of (5.4),
the second equation in (4.12) is obtained in the same way.

Specific solutions to system (3.4)
are more difficult to determine than in the case
of constant mean curvature, however, in what follows
a few examples are given. 

(i) As a simple example to begin with,
consider mean curvature $H$ given by the rational function
$$
H (z, \bar{z} ) = \frac{1}{1 + \lambda^2 ( z + \bar{z})^2},
\eqno(5.5)
$$
$\lambda$ is a real constant, and $H$ is real-valued
and analytic. By substituting $H$ into
(3.4), it can be checked that the function
$$
\rho = \lambda ( z + \bar{z} ) = \bar{\rho},
\eqno(5.6)
$$
is a solution to system (3.4). Moreover, $\partial \rho 
= \lambda = \bar{\partial} \bar{\rho}$ and the
corresponding solution to system (2.1) is given by
$$
\psi_1 = \epsilon \frac{\lambda^{3/2} (z + \bar{z})}
{ (1 + \lambda^2 ( z + \bar{z})^2)^{1/2}},
\qquad
\psi_2 = \epsilon \frac{\lambda^{1/2} }
{(1 + \lambda^2 ( z + \bar{z})^2)^{1/2}},
\quad
\epsilon = \pm 1.
\eqno(5.7)
$$
Then (2.2) induces a surface with mean curvature 
function given in (5.5).

(i) Let the mean curvature function be given 
as follows
$$
H ( z, \bar{z} ) = \frac{\exp ( \lambda ( z + \bar{z}))}
{1 + \exp ( \lambda ( z + \bar{z}))^2},
\eqno(5.8)
$$
where $\lambda$ is an arbitrary  real nonzero constant. 
In this case, $H$ is real valued and analytic. 
Substituting (5.8) into (3.4), it can be verified
that the function $\rho$ given by
$$
\rho = \exp ( \lambda ( z + \bar{z})) = \bar{\rho},
\eqno(5.9)
$$
is a solution to system (3.4). 
From (5.9), we calculate
$$
\partial \rho =  \lambda
\exp ( \lambda ( z + \bar{z})) = \bar{\partial} \bar{\rho}.
\eqno(5.10)
$$
and from (3.3), the functions $\psi_i$ are given by
$$
\psi_1 = \epsilon \exp ( \lambda ( z + \bar{z}))
\frac{\lambda^{1/2} }
{( 1 + \exp (2 \lambda ( z + \bar{z})))^{1/2}},
\qquad
\psi_2 = \epsilon \frac{\lambda^{1/2} }
{( 1 + \exp ( 2 \lambda ( z + \bar{z})))^{1/2}},
\quad
\epsilon = \pm 1.
\eqno(5.11)
$$
The $\psi_i$ in (5.11) can be used to determine
the coordinates of a surface by means of the
inducing prescription (2.2) corresponding to
the function $H$ in (5.8).

(iii) As a final example, let the mean curvature 
function be given by 
$$
H = A \tan ( A ( z + \bar{z})) 
\frac{\cos^2 (A (z + \bar{z})) +2 } 
{ \cos^2 ( A (z + \bar{z})) - 2},
\eqno(5.12)
$$
where $A$ is an arbitrary, real nonzero
constant, so that $H$ is real valued.
With (5.12) placed in (3.4), it is found that
$$
\rho = \sin ( A ( z + \bar{z})) = \bar{\rho},
\eqno(5.13)
$$
is a solution to system (3.4). From
(5.13), we have that
$$
\partial \rho = A \cos ( A (z + \bar{z})) =
\bar{\partial} \bar{\rho}.
$$
The corresponding solutions to (2.1) are then
$$
\psi_1 = \epsilon \sin (A (z + \bar{z}))
\frac{(A \cos ( A (z + \bar{z})))^{1/2}}{H^{1/2}
( 1 + \sin^2 ( A (z + \bar{z})))},
\quad
\psi_2 = \epsilon \frac{(A \cos ( A ( z + \bar{z})))^{1/2}}{H^{1/2}
( 1 + \sin^2 (A (z + \bar{z})))},
\quad
\epsilon = \pm 1.
\eqno(5.14)
$$
The coordinates of the surface follow by using (2.2)
with the $\psi_i$ given in these three  examples (i)-(iii).

{\bf VI. Summary and Physical Application.}

In this paper, an investigation of the
generalized Weierstrass system for the case in which the
mean curvature of the surface is not
constant has been initiated.
It has been shown, for example, that there
exists a transformation from solutions
of the generalized Weierstrass system (2.1) 
to solutions of a nonlinear sigma model
system (3.4), and of course, a transformation
from solutions of (3.4) to solutions of
system (2.1). We would like to conclude by
presenting a few remarks as to how the ideas described
here can be brought to bear on the study of string
theory.

A rigid string action can be written in the form $[15,16]$
$$
A = \gamma \int \int \, dS  +
\alpha \int \int H^2 \, d S,
$$
where $\alpha$ and $\gamma$ are some constants,
and integration takes place over the
string world surface $S$ which has extrinsic
mean curvature $H$. Moreover, one can
consider restricting ourselves to 
three-dimensional space-time and to a Euclidean version 
of the model. By varying the function $A$, very
simple equations relating fundamental
geometrical invariants of the string world surface
can be obtained. The Euler-Lagrange equation
following from the vanishing of the normal
variation of the rigid string action is
found to be given by
$$
- 2 \gamma H + \alpha ( \Delta H + 2 H^3 + R H) = 0,
$$
where $H$ is the mean curvature, $R$ is the
scalar curvature which is related to the Gauss
curvature by $K =- R/2$ and $\Delta$ is the
Laplace-Beltrami operator given on the surface $M$.

The theory described here then provides a
connection between the solution of an equation of 
motion and a method whereby the coordinate
expressions of an actual surface can be calculated
using (2.2), in a straightforward way.
It is hoped that further work will yield 
further results in higher dimensional spaces,
such as Minkowski space relatively soon.

\vspace{3mm}
{\bf References.}

\noindent
$[1]$ K. Weierstrass, Fortsetzung der Untersuchung
\"{u}ber die Minimalfl\"{a}chen, Mathematische Werke,
Vol. 3 (Verlagsbuchhandlung, Hillesheim, 1866) pp.
219-248; G. Darboux, Lecons sur Syst\`emes 
Orthogonaux et les Coordonnes Curvilignes
(Gauthier-Villars, Paris, 1910).  \\
$[2]$ B. Konopelchenko, ``Induced surfaces and their
integrable dynamics,'' Stud. Appl. Math. {\bf 96},
9-51 (1996).   \\
$[3]$ B. G. Konopelchenko and I. A. Taimanov, ``
Constant mean curvature surfaces via an integrable
dynamical system,'' J. Phys. {\bf A 29}, 1261-1265 (1996).  \\
$[4]$ P. Bracken, A. M. Grundland and L. Martina, ``
The Weierstrass-Enneper System for Constant Mean Curvature 
Surfaces and the Completely Integrable Sigma Model,''
J. Math. Phys. {\bf 40}, 3379-3402 (19999).  \\
$[5]$ P. Bracken and A. M. Grundland, ``Symmetry
Properties and Explicit Solutions of the 
Generalized Weierstrass System'', J. Math.
Phys. {\bf 42}, 1250-1282 (2001).  \\
$[6]$ P. Bracken and A. M. Grundland, ``On Certain
Classes of Solutions of the Weierstrass-Enneper 
System Inducing Constant Mean Curvature Surfaces'',
J. of Nonlinear Mathematical Physics, {\bf 6}, 294-313 (1999).  \\
$[7]$ P. Bracken and A. M. Grundland, ``On Complete
Integrability of the Generalized Weierstrass System'', J. of Nonlinear
Mathematical Physics, {\bf 9}, 229-247 (2002). \\
$[8]$ W. Zakrzewski, Low Dimensional Sigma-Models
(Hilger, New York, 1989).  \\
$[9]$ D. Nelson, T. Piran, and S. Weinberg,
Statistical Mechanics of Membranes and Surfaces
(World Scientific, Singapore, 1992).  \\
$[10]$ D. G. Gross, C. N. Pope, and S. Weinberg,
Two-dimensional Quantum Gravity and Random Surfaces
(World Scientific, Singapore, 1992).  \\
$[11]$ A. M. Grundland, L. Martina and
G. Rideau, Partial differential equations
with differential constraints, in CRM Proceedings
and Lecture Notes (American Mathematical Society,
Providence, RI, 1997), Vol. 11, pp. 135-154.  \\
$[12]$ V. G. Makhankov and O. K. Pashaev, 
Integrable Pseudospin Models in Condensed
Matter, Sov. Sci. Rev. Math. Phys. {\bf 9},
1-151 (1992).  \\
$[13]$ P. Bracken, ``Spin Model Equations, Connections
with Integrable Systems and Applications to 
Magnetic Vortices'', Int. J. of Mod. Physics,
{\bf B 17}, 4325-4537 (2003).  \\
$[14]$ P. Bracken, ``Reductions of Chern-Simons Theory
to Integrable Systems which have Geometric Applications'',
Int. J. Mod. Physics {\bf B 18}, 1261-1275 (2004). \\
$[15]$ A. Polyakov, ``Fine structure of strings'',
Nucl. Phys. {\bf B 286}, 406-412, (1986).  \\
$[16]$ H. Kleinert, ``The membrane properties of
condensing strings'', Phys. Letts {\bf B 174}, 
335-338, (1986).  \\

\end{document}